\documentclass[a4wide,12pt]{article}
\usepackage{hyperref}
\usepackage[margin=1.2in]{geometry}
\usepackage[utf8]{inputenc}
\usepackage [english] {babel} 
\usepackage{amsmath,amssymb,latexsym}
\usepackage{slashed}
\usepackage{graphicx}
\usepackage{epstopdf}
\usepackage{verbatim}
\usepackage[sorting=none, backend=bibtex]{biblatex}
\usepackage{pict2e}
\usepackage{textcomp}
\bibliography{ref}
\DeclareGraphicsExtensions{.png}

\def \sec{\begin{section}}
\def \esec{\end{section}}

\def \at {\biggl{\vert}}
\def \al {\alpha}

\def \la {\lambda}
\def \La {\Lambda}

\def \Om {\Omega}

\def \ep {\epsilon}

\def \pr {\partial}
\def \ra {\rightarrow}

\def \beq { \begin{equation}}
\def \eeq {\end{equation}}
\def \emf {e^{-\beta m_f}}
\def \emaf {e^{\beta m_{a}}}

\def \ta {\tilde a}
\def \tq {\tilde q}
\def \tt {\tilde t}

\newcommand\Li{\operatorname{Li}}

\def \l {\left(}
\def \r {\right)}

\def \ll {\langle}
\def \rr {\rangle}

\def \up {\Upsilon}
\def \bm {b^{-1}}
\def \sqt {\sqrt{qt}}
\def \gm {g^{-1}}
\def \hqc {\frac{Q_c}{2}} 

\begin{document}
\begin{titlepage}

\begin{center}
{  \Large \bf Condensates and instanton  - torus knot duality. Hidden Physics at UV scale.}
\end{center}
\vspace{1mm}

\begin{center}
{\large
   A.~Gorsky$^{\,2,3}$ and  A.~Milekhin$^{\,1,2,3}$\\ }
\vspace{3mm}
$^1$Institute of Theoretical and Experimental Physics, B.~Cheryomushkinskaya 25, Moscow 117218, Russia \\
$^2$Moscow Institute of Physics and Technology, Dolgoprudny 141700, Russia \\
$^3$ Institute for Information Transmission Problems of Russian Academy of Science, B.~Karetnyi 19, Moscow 127051, Russia\\
\vspace{1cm}
gorsky@itep.ru, milekhin@itep.ru
\end{center}

\vspace{1cm}

\begin{center}
{\large \bf Abstract}
\end{center}
We establish the duality between the torus knot superpolynomials 
or the Poincaré polynomials of the  Khovanov homology and
particular condensates in $\Omega$-deformed 5D supersymmetric QED compactified on a circle 
with 5d Chern-Simons(CS) term.
It is explicitly shown that $n$-instanton contribution 
to the condensate of the massless flavor
in the background of four-observable, exactly
coincides with the superpolynomial of the $T(n,nk+1)$ torus knot where $k$ - is the 
level of CS term. In contrast to the previously known results, the particular torus knot 
corresponds not to the partition function of the gauge theory but
to the particular instanton contribution and summation over the knots has to
be performed in order to obtain the complete answer.
The instantons are  sitting almost at the top of each other and the physics
of the "fat point" where the  UV degrees of freedom are slaved 
with point-like instantons  turns out to be quite rich. Also also see knot polynomials in the quantum mechanics on the instanton moduli space. We consider the different limits of this 
correspondence focusing at their physical interpretation and compare the algebraic structures
at the both sides of the correspondence. Using the AGT correspondence, we 
establish a connection between superpolynomials for unknots and q-deformed DOZZ factors.

\end{titlepage}

\tableofcontents

\newpage

\section{Introduction}

The knot invariants were introduced into the QFT framework long time ago \cite{witten} however
the subject has been getting new impact during the last decade. It turns out that the knot invariants should be considered
in the QFT in much more broader context. They are playing several interesting roles besides the original
interpretation as the Wilson loop observables in the CS theory. New approaches to their evaluation have been developed.
It was recognized in \cite{ov} that the 
open topological strings with Calabi-Yau target space provide an effective tool to derive the knot 
invariants and simultaneously knot invariants count the particular BPS states in the gauge theory. 
The target space for the topological string was identified with $\mathcal{O}(-1)\times \mathcal{O}(-1) \rightarrow CP_1$ and
the knot is selected by the Lagrangian brane  wrapping the Lagrangian submanifold. 
The boundary of the open string worldsheet is fixed at the Lagrangian submanifold in the CY internal space. 
The recent discussion on the topological string approach to the knot invariants can be found in \cite{einard,klemm}. 
The simplicity
of torus knots stimulated the derivation of very explicit results and representations
for them\cite{labastida,ov}.

The progress in the knot theory brings on the scene the Khovanov-Rozansky homologies which categorize
the HOMFLY polynomial. The Poincaré polynomial of the Khovanov-Rozansky homologies has been 
interpreted in the framework of the topological strings in \cite{gsv} and it was shown that
such Poincaré polynomial, called superpolynomial \cite{dgr}, provides the refined counting of the
BPS states. The Khovanov-Rozansky homologies were also related with the space of the solutions to the
topological fields theories in four and five dimensions \cite{wittennew,wittengai}. The way to evaluate 
the superpolynomials for some class of knots has been suggested in \cite{agash} via the refined 
Chern-Simons theory or equivalently the particular matrix model.

Another way the knot invariants are related with the gauge theories concerns the 3d/3d duality
\cite{ggd} which relates the 3d theory on the submanifold in the CY space and the 3d 
SUSY gauge theory. The knot complement yields the particular 3d SUSY gauge theory with some matter content.
The superpolynomial is related to the partition sum of 3d theory and the parameters $(A,q,t)$ in the superpolynomial were
identified with the mass and two equivariant parameters with respect to two independent rotations
in $\mathbb{R}^4$ \cite{gukov12}. The relation between the partition function on the vortex and the knot polynomials has been discussed 
in \cite{holland}. If we introduce the defects, say 2d defect in 4d theory or
3d defect in 5d theory, any physical phenomena should be recognized equivalently from the worldvolume theories
of all branes involved into configurations. This simple
argument suggested long time ago \cite{dorey} works well and provide some interesting
crosschecks (see, for instance \cite{triality}). In particular, all knot invariants should be recognized
by all participants of the configuration.

There is one more important characteristic of the knot -- so-called A-polynomial and its generalization --
super-A-polynomial \cite{gukov12} depending on the set of variables $(x,y)$ 
which become operators upon the quantization of the $(x,y)$ symplectic pair. The A-polynomial
defines the twisted superpotential in 3d theory \cite{gukov12,av}. 
The simplest interpretation of the $(x,y)$ variables concerns the realization of the 3d theory as the
theory on the domain wall separating two 4d theories \cite{yamazaki, gukovwall}. They
are  identified with the Wilson and 't Hooft loops variables. The nice review on the subject
can be found in \cite{gukovrev}.

There was some parallel progress in mathematics concerning the homologies of the torus knots and links. In what follows
we shall use quite recent  results concerning superpolynomials of the torus knots and their relation
with the higher $(q,t)$-Catalan numbers \cite{ors,gors, gorsky10, gorneg}.

Another interesting line of development motivating our consideration concerns the UV completion of the different theories and the 
features of the decoupling of the heavy degrees of freedom. Some  phenomena can happen and we know from the textbooks 
that the heavy degrees of freedom decouple when the masses of the corresponding excitations get large enough with the 
only exception - 
anomaly which matters at any scale since it arises from the spectral flow. However one could wonder what happens at 
the non-perturbative level.  
The issue of the role of small size instantons in the RG was raised long time ago in QCD when the integrals over the instanton size tend to diverge. 
The issue of the point-like instantons is also very important in the considerations of the so-called contact terms which measure the 
difference between the products of 
the different observables in UV and IR regions (see in this context, for instance \cite{lns97,moore}).
Therefore some regularization is needed to handle with this region in the instanton moduli space.

There were several attempts to analyze the point-like instantons carefully imposing a kind of regularization. 
We can mention the freckled instantons in \cite{lns}, Abelian instantons in the non-commutative gauge theories \cite{ns98} and the Abelian instantons on 
the commutative $\mathbb{R}^4$ blow-upped in a few points \cite{braden}. In all these cases one could define the corresponding 
solutions to the equations of motion with the non-vanishing topological charges. The last example concerns 
the Abelian instantons in the $\Omega$-deformed abelian theory where the point-like instantons can be defined
as well \cite{nekrasov14}. The different deformations provide the possibility to work with the point-like instantons
in a well-defined manner. In these cases we can pose the question concerning the role of these defects in the
non-perturbative RG flows.

Moreover one could ask what is the fate of the extended non=perturbative
configurations which involve the heavy degrees of freedom in a non-trivial way. This issue has been examined in
\cite{rastelli,holland, razamat} where the vortex solution involving the "heavy fields" was considered. 
The starting point is the superconformal theory then one adds the bi-fundamental matter. One considers the non-Abelian string in the
emerging UV theory. At the next step the RG flow generated by the FI term has been analyzed and it was argued
that the remnant of the UV theory is the surface operator, that is, the non-Abelian string with the infinite tension. 
Hence the decoupling in the non-perturbative sector for the extended object is incomplete - we obtain at the end the  defect with the infinite tension which 
provides the boundary conditions for the fields.

The third motivation for this study has more physical origin. When investigating the superfluidity it is very useful to 
rotate the system since the 
superfluid component of the current can be extracted in this way. In the first quantization approach 
the density of the superfluid component is related to the
 correlator of the winding numbers. The $\Omega$ deformation which is 
essentially two independent rotations in $\mathbb{R}^4$ was introduced by Nekrasov to regularize the 
instanton moduli space. On the other hand, it allows to look at  the response of the ground state of to 
the rotations like in superfluidity. Since the curvature of the 
graviphoton field is just the angular velocity we could consider the 
behavior of the partition function at small angular velocities $\epsilon_1,\epsilon_2$. 
It turns out that the dependence on the angular velocities  is very simple \cite{nekrasov} and the 
derivative of the partition function with respect to the angular velocity 
yields the  mean angular momentum of the system \cite{gorskygravi}. It can be seen 
immediately that there is non-vanishing density of the angular momentum and 
one could be interested in its origin. It is not a simple question what is the 
elementary rotator like the roton in the superfluid case in the 5d gauge theory. 
However no  doubts it should be identified as some non-perturbative configuration related to instanton.

The starting point of our analysis is the observation
made in  \cite{bgn14} concerning the relationship between certain knots 
invariants and the 5d SUSY gauge theory in the $\Omega$-background with CS term at level one and 
matter in fundamental. It was shown that the particular correlator in $U(1)$ SQED  
coincides with the sum over the bottom rows of the superpolynomials of 
the $T_{n,n+1}$ torus knot. Contrary to the previous relations 
between knots and the gauge theories in this case the particular gauge theory 
involves the infinite sum over the torus knots. 

In this paper we generalize the observation made in \cite{bgn14} and find the similar relation between the sums over the
torus  knot superpolynomials and the $U(1)$ 5d gauge theories with CS term. 
In more general situation the second derivative of Nekrasov partition function for 
5d SQED with CS term serves as the generating function for the 
superpolynomial of the torus knots. It is useful to interpret the 4-observable considered
in \cite{bgn14} in a bit different manner. We start with the superconformal 5d theory and add matter in fundamental.
Then the "Wilson loop" operator in \cite{bgn14}  can be considered as the derivative with respect to the mass of the 
one-loop determinant  of the matter in the fundamental at the infinite mass limit. To some extend the approach used 
in this paper is along the lines of development elaborated in \cite{wittennew,wittengai} where the knot homologies were
interpreted in terms of the particular solutions to the equations of motion in 5D SUSY gauge theories. 
However from our analysis it is clear that the proper generating function for the torus knot superpolynomials implies the particular matter content in the
5D theory.

One more lesson concerns the question about the mutual back reaction of IR and UV degrees of freedom. 
Our answer for the torus knot superpotential  involves the derivatives of the partition function in 5D theory with respect to the masses of the 
light and the "regulator" flavors hence
it allows the twofold interpretation. First, it can be treated 
as a kind of the point-like instanton renormalization of the VEV $\frac{d}{dM} \ll {\tilde \psi} \psi \rr$ of the light flavor in the $\Omega$-background and is 
treated as the back reaction of the UV degrees of freedom on the 
condensate of the massless flavor. The operator has non-vanishing anomalous dimension hence to some extend the superpolynomial yields the instanton contribution to the 
anomalous dimensions of the composite operator. Oppositely the same correlator can be read in the opposite order and can be
thought of as a kind of the backreaction of the light flavor on the defect which involves the UV degrees of freedom.

Summarizing,  we shall demonstrate that the gauge theory
whose $n$-instanton contribution to the particular correlator  coincides with the 
superpolynomial of $T_{n,nk+1}$ torus knot is the  $U(1)$ 5d gauge theory with one 
compact dimension,
CS term at level $k$, 2 flavors in fundamental and one flavor in the 
anti-fundamental representation. 
One mass of the fundamental tends to zero while the second tends to infinity. 
The mass of the anti-fundamental is arbitrary. The remnant 
of the heavy flavor in the IR is the chiral ring operator.

The paper is organized as follows. In Section \ref{sec:sqed} we briefly remind the main facts 
concerning the 5d SQED with CS term
and focus at the decoupling procedure in this theory. In Section \ref{sec:qt} 
we describe the relation between the instanton contribution
to the derivative of condensate of the light hyper with respect to the regulator scale and the torus knot supepolynomials.
In Section \ref{sec:int} we attempt to interpret the result of calculation in terms
 of a kind of composite defect involving UV degrees of
freedom.  Section \ref{sec:lim} is devoted to the consideration of the different limits for parameters involved in the picture.
The interpretation of the correlator from the AGT dual Liouville theory viewpoint will be considered in Section \ref{sec:agt}. 
The question concerning the identification of the knot polynomials in the quantum mechanics on the instanton moduli space 
will be analyzed in Section \ref{sec:qm}. Our findings and the lines for the further developments are summarized in the Conclusion.

\section{Supersymmetric QED with CS term}
\label{sec:sqed}
\subsection{Fields, couplings, symmetries and Lagrangian with $\Omega$ deformation}
Five-dimensional supersymmetric QED involves of vector field $A_A$, four-component Dirac spinor $\lambda$ and Higgs field $\phi$, all lying in the adjoint representation of $U(1)$.
The Lagrangian reads as follows:
\beq
\mathcal{L}= -\frac{1}{4g^2} F_{AB} F^{AB} + \frac{1}{g^2}(\pr_A \phi)^2 + \cfrac{1}{g^2} \bar \lambda \gamma^A \pr_A \lambda  
\eeq
$\gamma^A, A=1,...,5$ are five-dimensional gamma matrices. 

When looking for the superfluidity it is useful to rotate the system to feel the dissipationless
component of the liquid. The same trick was used  by Nekrasov \cite{nekrasov}  when introducing the 
$\Omega$-background which corresponds to switching on the graviphoton field whose components
of curvature are identified with two independent angular velocities in $R^4$. The response of the
partition function in the $\Omega$-deformed theory yields the gravimagnetization of the ground state
or the average angular momentum of the system. To some extend it measures the " superfluid"
component of the vacuum state of the 4d gauge theory.

Let us start from the discussion of pure gauge $\mathcal N=2$ super Yang-Mills theory in presence of $\Omega$-background in four Euclidean dimensions. Then we will lift the theory to five-dimensions. 
The field content of the theory is the gauge field $A_m$, the complex scalar $\varphi, \bar{\varphi}$ and Weyl fermions $\Lambda^I_\alpha, \bar \Lambda^I_{\dot{\alpha}}$  in the adjoint of the $U(1)$ group. Here $m=1,\dots,4$, $I=1,2$ are $SU(2)_I$ R-symmetry index, $\alpha,\dot\alpha$ are the $SU(2)_L\times SU(2)_R$ spinor indices. To introduce $\Omega$-background one can consider a nontrivial fibration of $\mathbb R^4$ over a torus $T^2$ \cite{nekrasov},\cite{NekOk}. The six-dimensional metric is:

\begin{equation}
    ds^2=2dzd\bar z+\left( dx^m+\Omega^md\bar z+\bar\Omega^mdz \right)^2,
    \label{metric_Omega}
\end{equation}
where $(z,\bar z)$ are the complex coordinates on the torus and the four-dimensional vector $\Omega^m$ is defined as:

\begin{equation}
    \Omega^m=\Omega^{mn}x_n,\qquad \Omega^{mn}=\frac{1}{2\sqrt{2}}\begin{pmatrix}0&i\epsilon_1&0&0\\-i\epsilon_1&0&0&0\\0&0&0&-i\epsilon_2\\0&0&i\epsilon_2&0\end{pmatrix}.
     \label{omega}
\end{equation}

In general if $\Omega^{mn}$ is not (anti-)self-dual the supersymmetry in the deformed theory is broken. However one can insert R-symmetry Wilson loops to restore some supersymmetry \cite{NekOk}:

\begin{equation}
    A^I_J=-\frac12\Omega_{mn}\left( \bar\sigma^{mn} \right)^I_J d\bar z-\frac12\bar\Omega_{mn}\left( \bar\sigma^{mn} \right)^I_J dz.
    \label{Wilson}
\end{equation}

The most compact way to write down the supersymmetry transformations and the Lagrangian for the $\Omega$-deformed theory is to introduce 'long' scalars
(do not confuse them with $\mathcal{N}=1$ superfields):

\begin{equation}
    \Phi=\varphi+i\Omega^mD_m, \qquad \bar\Phi=\bar\varphi+i\bar\Omega^mD^m,
      \label{phi_def}
    \end{equation}

Then bosonic sector of the deformed Lagrangian reads as:
\begin{equation}
\begin{split}
  \mathcal L_\Omega=-\frac{1}{4g^2}F_{mn}F^{mn}+\frac{1}{g^2} D_m\Phi D^m\bar\Phi+\frac{1}{2g^2}\left[ \Phi,\bar\Phi \right]^2 = \\
  -\frac{1}{4g^2}F_{mn}F^{mn}+\cfrac{1}{g^2}(\pr_m \phi + F_{mn} \Omega^n)(\pr^m \phi - F^{mn} \Omega^n)+
  \frac{1}{2g^2}(i \Omega^m \pr_m \bar \phi + i \Om^m \pr_m \phi)^2
  \label{lagrangian}
\end{split}
\end{equation}

We can couple this theory to fundamental hypermultiplet, which consists of two scalars $q$, $\tilde q$ and two Weyl fermions $\psi$ and $\tilde \psi$ and 
characterized by two masses: $m$ and $\tilde m$, since $\mathcal{N}=2$ hypermultiplet is build from two $\mathcal{N}=1$ hypermultiplets with opposite
charges. Now the bosonic part reads as:
\beq
\begin{split}
\mathcal{L}_m=-\frac{1}{4g^2}F_{mn}F^{mn}+\cfrac{1}{g^2}(\pr_m \phi + F_{mn} \Omega^n)(\pr^m \phi - F^{mn} \Omega^n)+ \\
   \cfrac{1}{2} |D_m q|^2 + \cfrac{1}{2} |D_m \tilde q|^2 + \cfrac{2}{g^2}(i \pr_m(\Omega^m \bar \phi+ \Omega^m \phi)+g^2(\bar q q-\bar \tilde q \tilde q) )^2 + \\
 \frac{1}{2}|(\phi-m-i \Omega^m D_m)q|^2 + \frac{1}{2}|(\phi- \tilde m-i \Omega^m D_m)\tilde q|^2 + 2g^2|\tilde q q|^2
\end{split}
\eeq


General $\Omega$-deformation preserves only one supersymmetry\cite{NekOk}. It is convenient to introduce topological twist\cite{nekrasov} and take $SU_L(2)$ times diagonal subgroup of 
$SU_R(2) \times SU_I(2)$ to be Lorentz group. Then $\Lambda^I_{\dot \alpha}$ becomes scalar $\eta$ and self-dual tensor $\chi_{IJ}$, $\Lambda^I_\alpha$ becomes 
vector $\psi_I$, and $\psi, \bar \psi$ becomes $\theta, \nu_m, \omega_{mn}$. Supercharges have
similar fate. The scalar supercharge $Q=Q_\Om$ stays unbroken.


\subsection{On Decoupling procedure}

Decoupling of the heavy flavor in the 5d gauge theory is very delicate issue mainly due to the UV incompleteness of the theory. It was
discussed in many studies that the naive field theory intuition fails and the purely stringy
degrees of freedom like different D-branes emerge in the UV completion problem. It can be recognized in the different
ways, for instance, from the viewpoint of the ADHM quantum mechanics which describes the UV physics from the
viewpoint of the instanton particles. The ADHM quantum mechanics in this case involves the tiny issues at the threshold
when the continuum spectrum opens. It was assumed that in this quantum mechanics the stringy degrees of freedom
get manifested in the index calculations.

One more pattern of the nontrivial decoupling of the heavy degrees of freedom is provided by the 4d example
of the decoupling of the heavy flavor \cite{rastelli}. The naive decoupling of the heavy flavor fails and one finds
himself with the remnant surface operator supplemented by the operator acting in the flavor fugacity space. This operator
was identified with the integrable Hamiltonian of the Calogero-Ruijsenaars type \cite{rastelli}. 

In our paper we shall meet the subtleties with the UV completion as well. W the e start with the theory with the heavy flavor 
and try to decouple it. During this process we get the particular 4-observable as the remnant which seems to be
identified naturally with the domain wall in the $\Omega$-deformed 5d SQED. This is to some extend analogous to the 4d case 
however the 4-observable emerges instead of 2-observable remnant. We shall also see how the information about the UV completion can be extracted from 
the ADHM quantum mechanics on the instanton moduli space. The knot invariants encodes the particular set of states near threshold.

On the quantitative level we shall get the following remnant of the heavy flavor in the following way.
Although  we start with  three matter hypermultiplets, actually we need only two of them  since one has infinite mass.
 Now we will show that
the only effect from this heavy  hypermultiplet is an insertion of the operator $O=\int d^5x \exp(-\beta( \phi+ i A_5)$, 
where $\phi$ is a vector multiplet scalar and $A_5$ is the fifth component of the gauge field:
\beq
\label{def_o}
\lim_{m_2 \ra \infty} \cfrac{\exp(\beta m_2)}{\beta} \cfrac{\pr}{\pr m_2} Z^{U(1)}(m_1,m_2,m_3) = \ll O \rr_{m_1, m_3}^{U(1)}
\eeq
Note that upon reduction to four dimensions $\phi+i A_5$ becomes complex scalar Higgs field. Suppose for a while that we 
are considering four-dimensional theory without $\Om$-deformation. Then integrating out heavy hypermultiplet will produce usual Coleman-Weinberg 
potential
\beq
(\phi+m_2)^2  \l \log \l \cfrac{\phi+m_2}{\La_{UV}} \r -1 \r =\int_0^{\infty} \cfrac{dt}{t^3} \exp(-t(\phi+m_2))
\eeq
In order to lift this expression to a five-dimensional theory\cite{Nek5d}, we have to sum over the Kaluza-Klein modes, that is, add 
$\cfrac{2 \pi i n}{R} + i A_5$ to
$\phi+m_2$ and sum over $n$. This will result in
\beq
\Li_3 \l e^{-2 \pi R(\phi+m_2+i A_5)} \r
\eeq
Which is for large $m_2$ is just $\exp(-2 \pi R(\phi+A_5+m_2))$
So we have reproduced eq. (\ref{def_o}) with the operator $O=\int d^5x \exp \l -\beta \l \phi(x)+ i A_5 \r  \r$. 

When switch to the $\Om$-deformed theory, almost all supersymmetries are broken and the chiral ring gets deformed, since
conventional Higgs scalar is not equivariantly closed:
\beq
Q_\Om \phi = \Om^\mu A_\mu
\eeq
 Appropriate deformation of
complex Higgs field $\phi$ in four-dimensions such that
\beq
Q_\Om \Phi=0
\eeq
was build in \cite{NekOk} and we 
claim that the $\exp(-\beta \Phi)$ with deformed $\Phi$, is exactly the operator we 
need even in the Omega-deformed theory. In the next section we will demonstrate this statement by a direct computation.

\section{Superpolynomial of torus knots and 5d SQED }
\label{sec:qt}

In this Section we shall explore the localization formulas for the instanton Nekrasov-like partition sums
in the 5d SUSY QED. Therefore we look for the proper physical theory which would involves the knot invariants
in a rational clear-cut manner.
In this paper we extend the proposal formulated in \cite{bgn14}, which relates $q,t$-Catalan numbers represented  
the bottom row of the superpolynomials of $T_{n,n+1}$ torus knots.

We shall evaluate the K-theoric equivariant integral over the moduli space of the instantons. It is equal to equivariant Euler characteristic of the tautological line bundle $V$ over the 
Hilbert scheme:
\beq
C_n(q,t)=\chi^T(Hilb^n(\mathbb{C}^2),\Lambda^n V)
\eeq
where $q,t$ are equivariant parameters for the natural torus $T$ action on $\mathbb{C}^2$. $C_n(q,t)$ are called q,t-Catalan numbers.

First, recall some relevant mathematical results.

In \cite{qt_cat} Haiman and Garsia introduced the following generalization of Catalan numbers:
\begin{equation}
 C_{n}(q,t)=\sum_{\la : |\la|=n } \large{ \cfrac{t^{2 \sum l} q^{2 \sum a}(1-t)(1-q)\prod^{0,0}(1-q^{a'}t^{l'})(\sum q^{a'} t^{l'})}
{\prod(q^a-t^{l+1})\prod(t^l-q^{a+1})} }
\end{equation}
where all sums and products are taken over partition $\la$. $l$ and $a$ denote leg and arm, whereas $l'$ and $a'$ denote coleg and coarm respectively. 
$\prod^{0,0}$ denote the omission of $(0,0)$ box. 
In case $q=t=1$, $C_n(1,1)=\cfrac{1}{n+1} \left(
\begin{array}{c}
2n\\
n\\
\end{array}
\right)$

It is also useful to present the expressions for the so-called higher Catalan numbers introduced in \cite{haiman}. They
can be represented in terms of the Young diagrams as follows 
\beq
C_n^k(q,t)=\sum_{\la : |\la|=n } \large{ \cfrac{t^{(k+1)\sum l} q^{(k+1) \sum a}(1-t)(1-q)\prod^{0,0}(1-q^{a'}t^{l'})(\sum q^{a'} t^{l'})}
{\prod(q^a-t^{l+1})\prod(t^l-q^{a+1})} }
\eeq
In what follows we shall identify the index $k$ with the level of 5d Chern-Simons term. The shift $k\rightarrow k+1$ corresponds
to the decoupling of one flavor in the 5d supersymmetric SQED.

In \cite{gorsky10} it was shown that these numbers calculate Poincaré polynomial for a plain curve singularity corresponding to $(n+1,n)$ torus knot.
Furthermore, in \cite{ors} was conjectured the following expression for a superpolynomial for $(nk+1,n)$ torus knot:
\begin{eqnarray}
P(A,q,t)_{nk+1,n}= \\ \nonumber
 \sum_{\la : |\la|=n } \large{ \cfrac{t^{(k+1) \sum l} q^{(k+1) \sum a}(1-t)(1-q)\prod^{0,0}(1+A q^{-a'}t^{-l'})\prod^{0,0}(1-q^{a'}t^{l'})
(\sum q^{a'} t^{l'})}{\prod(q^a-t^{l+1})\prod(t^l-q^{a+1})} }
\end{eqnarray}

In this paper we extend the proposal formulated in \cite{bgn14}, which relates $q,t$-Catalan numbers and certain vacuum expectation value in
five-dimensional $U(1)$ gauge theory in the $\Om$-deformation. We claim that the above superpolynomial could be obtained via five-dimensional $U(1)$ gauge 
theory with 2 fundamental flavors with masses $m_f,M$, one anti-fundamental flavor with the mass $m_a$ and Chern-Simons term with the coupling $k$:
\begin{equation}
 P(A,q,t)_{n,nk+1}= t^{-n/2} q^{-n/2} \cfrac{1}{1+A} \cfrac{\exp(\beta M)}{\beta^2} \cfrac{\pr}{\pr m_f}\cfrac{\pr}{\pr M}  Z^{U(1)}_n(m_f,m_a,M) \at_{m_f = 0,\ M \ra \infty} 
\end{equation}
where $Z^{U(1)}_n$ is $n$-instanton contribution to the partition function. 

\textbf{NB:} our choice of variables is different from one adopted in \cite{ors}. We will perform the identification of variables when we
discuss various limits of these formulas.

\begin{figure}[h]
\begin{center}
\setlength{\unitlength}{1.4cm}
\begin{picture}(6,4)
\linethickness{0.3mm}
\put(3,2){\line(1,0){1}}
\put(4,2){\line(0,1){1}}
\put(3,2){\line(0,-1){1}}
\put(4,2){\line(1,-1){1}}
\put(5,1){\line(1,0){1}}
\put(5,1){\line(0,-1){1}}
\put(2,3){\line(1,-1){1}}
\put(1,3){\line(1,0){1}}
\put(2,3){\line(0,1){1}}
\put(3.2,2.1){\makebox{Q, $\lambda$}}
\put(1.5,2.3){\makebox{$Q_f$, $\mu_1$}}
\put(4.6,1.5){\makebox{$Q_a$, $\mu_2$}}
\end{picture}
\caption{$\mathcal{O}(-1) \times \mathcal{O}(-1) \ra \mathbb{P}^1$ with two blow-ups corresponding
to the 5D SQED with two flavors and zero CS term}
\label{fg:rc}
\end{center}
\end{figure}
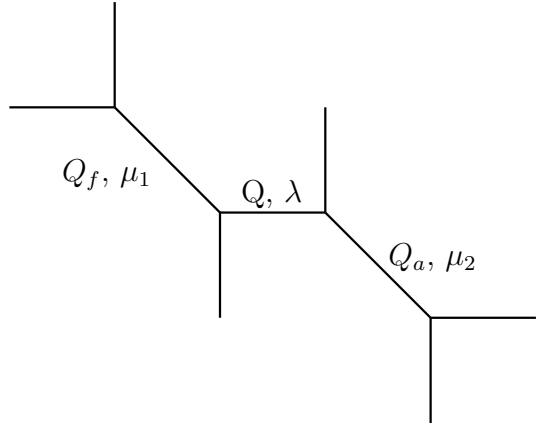

 One of the dimensions is compactified on a circle with radius $\beta$. We denote the $\Om$-background parameters by $\ep_1$ and $\ep_2$. Then 
\begin{eqnarray}
 t=\exp(-\beta \ep_1) \\
 q=\exp(-\beta \ep_2) \\
 A=-\exp(\beta m_a) 
\end{eqnarray}
We are going to prove this relation using the refined topological vertex technique\cite{vafa07}. According to \cite{iqbal08}, the full partition
function in the case of one fundamental flavor and one anti-fundamental flavor is given by:\footnote{note that in our notations $t \ra 1/t$}:
\beq
 Z^{U(1)}(m_f,m_a) =\large{ \sum_\la (-Q)^{|\la|} t^{|\la|/2} q^{|\la|/2} \times} \nonumber
\eeq
\beq
\large{  \dfrac{ t^{\sum l} q^{\sum a} \prod_{i=1,j=1}^{\infty} 
(1-Q_f q^{i-1/2} t^{\la_i-j+1/2}) (1-Q_a q^{-\la_i^t + j-1/2} t^{1/2-i}) }{\prod(t^l-q^{a+1})(t^{l+1}-q^a)} }
\eeq
where K\"ahler parameters: $Q_f= \exp(-\beta m_f)/\sqrt{q t}, Q_a = \sqt \exp(\beta m_a) $, $Q$ 
defines the coupling constant via $Q=\exp(-\beta/g)$. Corresponding three-dimensional Calabi-Yau geometry is represented on the Figure \ref{fg:rc}. 
Perturbative part is given by:
\begin{equation}
 Z^{U(1),pert}(m_f,m_a) = \prod_{i=1,j=1}^{\infty} (1-Q_f q^{i-1/2} t^{-j+1/2}) (1-Q_a q^{j-1/2} t^{1/2-i})
\end{equation}
Then n-instanton contribution is given by:
\beq
 Z^{U(1)}_n(m_f,m_a) =\large{ \sum_{|\la|=n} (-Q)^{|\la|} t^{|\la|/2} q^{|\la|/2} \times} \nonumber
\eeq
\beq
\large{ \dfrac{ t^{\sum l} q^{\sum a} \prod 
(1-\exp(-\beta m_f) t^{l'} q^{a'}) (1-\exp(\beta m_a) t^{-l'} q^{-a'}) }{\prod(t^l-q^{a+1})(t^{l+1}-q^a)} }
\eeq
Factors like 
\begin{equation}
\prod (1-\exp(-\beta m_f) t^{l'} q^{a'})  
\end{equation}
correspond to chiral matter contribution. $U(1)$ gauge part contributes:
\begin{equation}
 \large{ \sum_{|\la|=n} (-Q)^{|\la|} t^{|\la|/2} q^{|\la|/2} \dfrac{ t^{\sum l} q^{\sum a} }{\prod(t^l-q^{a+1})(t^{l+1}-q^a)}}
\end{equation}

Therefore, for the superpolynomial we need:

\begin{itemize}

\item In order to obtain a
factor $\prod^{0,0}(1-q^{a'}t^{l'})$ in the superpolynomial we have add a zero mass chiral multiplet and differentiate with respect to its mass.

\item To obtain $\sum q^{a'} t^{l'}$ we take another chiral multiplet, differentiate with respect to its mass and after that we send the mass to infinity.

\item Factor $\prod^{0,0}(1+A q^{-a'}t^{-l'})$ comes from the anti-fundamental multiplet.

\item Finally, it is well-known that the Chern-Simons action with the 
coupling constant $k$ contributes $t^{k \sum l} q^{k \sum a}$ - it can be easily seen in the above formulas if we remember that the Cherm-Simons term emerges as
a one-loop effect from $k$ very heavy chiral multiplets. However  Chern-Simons coupling affects the whole geometry. The coupling is actually
given by the intersection number of two-cycles on the manifold\cite{morrison,klemm08}. For example, the theory with coupling $k=1$ and without flavors is given by 
geometry $\mathcal{O}(0) \times \mathcal{O}(-2) \ra \mathbb{P}^1$ - see Figure \ref{fg:k1}. Recall, that the $n$-instanton contribution arises from the worldsheet instanton wrapping the 
base $\mathbb{P}^1$ $n$ times. This fact suggests that the shift $1 \ra kn+1$ in the torus knot
 is actually an analogue of the Witten effect, since the instanton wraps around the other two-cycle $kn$ additional 
times and acquires additional charge. 
\end{itemize}
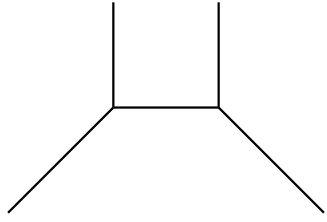
\begin{figure}
\begin{center}
\setlength{\unitlength}{1.4cm}
\begin{picture}(3,3)
\linethickness{0.3mm}
\put(0,0){\line(1,1){1}}
\put(1,1){\line(0,1){1}}
\put(1,1){\line(1,0){1}}
\put(2,1){\line(0,1){1}}
\put(2,1){\line(1,-1){1}}
\end{picture}
\end{center}
\begin{center}
\caption{$\mathcal{O}(0) \times \mathcal{O}(-2) \ra \mathbb{P}^1$ geometry corresponding to 5D SQED with the 
Chern-Simons coupling $k=1$, but without any flavors}
\label{fg:k1}
\end{center}
\end{figure}
Now let us return to the operator $\exp(-\beta \Phi)$.
Consider the term $(-1)(1-t)(1-q) \sum q^{a'} t^{l'}$ in the original expression for a superpolynomial. If $\la_i$ is the length of i-th row, then 
\beq
(1-t)\sum_{i=1}(q^{\la_i} t^{i-1} - t^{i-1}) = \sum_{i=1}( q^{\la_i} t^{i-1} - t^{i-1} - q^{\la_i} t^{i} + t^{i})
\eeq 
where the sum is over rows in a particular Young diagram $\la$. The last expression is exactly what we will obtain if we calculate the vev of $\exp(-\beta \phi)$
using Nekrasov formulas(\cite{NekOk} eq. (4.19)) - in this approach one introduces the profile function:
\beq
f_{\la,\ep_1,\ep_2}(x)= |x| + \sum_{i=1} \l |x+\ep_1 - \ep_2 \la_i  - \ep_1 i| - |x-\ep_2 \la_i - \ep_1 i|-|x+\ep_1-\ep_1 i|+|x- \ep_1 i| \r
\eeq
Then contribution to the vev of $\phi^n$ is given by
\beq
\cfrac{1}{2} \int_{-\infty}^{+\infty} dx \  x^n f^{''}_{\la,\ep_1,\ep_2}(x)
\eeq
We see that even in the presence of Omega-deformation decoupling of the heavy flavor leads to the insertion of $\exp(-\beta \phi)$.

Now we can evaluate the whole partition function summing over the instanton contributions
\beq
Z^{Nek}(m,\epsilon_1,\epsilon_2, \beta, Q )= \sum_n (-Q)^n Z_n(m,\epsilon_1,\epsilon_2, \beta)
\eeq
Where the whole partition function obeys some interesting equations as a function of its arguments. In the NS limit when
$\epsilon_2=0 \ra t=1$  the summation of q-Catalan numbers can be performed explicitly and yields \cite{bg14}:
\beq
P(q,Q)= \cfrac{\exp(\beta M)}{\beta^2} \cfrac{\pr}{\pr m_f}\cfrac{\pr}{\pr M} Z^{Nek}(m_f,M,q, Q/\sqrt{q} ) \at_{\ m_f = 0,\ M \ra \infty}  =\frac{A_q(Q q^2)}{A_{q}(Q q)}
\eeq
where $A_q(s)$ is the q-Airy function:
\beq
A_q(s) = \sum_k \cfrac{s^k q^{k^2}}{(q;q)_k}
\eeq
where $(z;q)_k=\prod_{l=0}^{k-1}(1-z q^l)$ is Pochhammer symbol. This implies that the condensate obeys the following relation:
\beq
\label{airy}
P(q,Q)=1-Q P(q,Q) P(q,qQ)
\eeq
Unfortunately, we do not know any field-theoretic explanation of this relation. We will return to this question when we will be discussing the stable limit $k \ra \infty$.

\section{The Attempt of interpretation}
\label{sec:int} 
\subsection{Point-like Abelian instantons }

In this Section we shall consider the physical picture behind the duality found. 
It implies that we have $n$ point-like instantons sitting almost at the top of each other
in the background provided by the nonlocal operator $\exp(- \beta {\Phi})$. When the $\Omega$
deformation is switched off the operator becomes local therefore the physical picture we shall
try to develop should respect this property. Another suggesting argument goes as follows. Consider the 
limit of $\epsilon_1,\epsilon_2 \rightarrow 0$ when the Nekrasov partition function is reduced to the form
\beq
Z_{Nek}\propto \exp \l \frac{F}{\epsilon_1 \epsilon_2} \r
\eeq
Having in mind that $\epsilon_1,\epsilon_2 $ are two angular velocities the simple argument shows that there is the
average angular momentum  $<J> \neq 0$ in the system \cite{gorskygravi} and one could say about the 
gravimagnetization of the ground state. Combining these arguments we could suspect that the microscopic state we are dealing with is built  
from the regulator degree of freedom, is nonlocal, has instanton charge $n$ and  some angular momentum.
This is the qualitative  description of the part of this extended object in $R^4\times S^1$.

Note that a somewhat similar situation occurs in the description of the 
nonperturbative effects in the ABJM model \cite{marino} where the membrane M2 instantons wrapping the (m,n) cycle in the internal space yields the corresponding contribution
to the partition function at strong coupling regime. 
The brane interpretation of the nonpeturbative effects at weak coupling is not completely clarified in that case.

We have to combine two parts into the worldvolume of some brane. 
First of all let us comment what are natural configurations in D=5 which obey the required property. The first candidate
is the dyonic instanton \cite{tong} or its supergravity counterpart - supertube. The dyonic instanton has the instanton charge Q , F1 charge P and D2 dipole charge/ It has the geometry of the cylinder with the distributed charge densities  and its angular momentum is proportional to the product of two charges $J\propto PQ$. In the other duality frame it is presented by the D3 brane with the KK momentum \cite{hashimoto}. In this case the defect could be represented by the M5 brane supplemented by the instanton charges.

The second candidate is the D6-D0 state which corresponds to the rotating black hole \cite{kerr}. This configuration
can be BPS in some region of parameters \cite{wittend6}. From the field theory viewpoint it represents the 
domain wall configuration in D=5 gauge theory which carries the additional angular momentum. In the 4d dimensional 
$\Omega$ -deformed gauge theory such closed domain wall does exist \cite{bg14} and has the geometry of the 
squashed sphere $S^3_b$ where $b^2=\frac{\epsilon_1}{\epsilon_2}$. Therefore the candidate defect would
have the worldsheet $S^3_b\times S^1\times M$ in this case.

In all cases we assume that the key contribution for the mechanism preventing the closed object from shrinking
is the angular momentum. When the SUSY is broken in some way the additional source come from the 
difference between the energies providing the stabilizing pressure.
It is natural to expect that the nonperturbative configuration is sensitive to the $\Omega$ background and moreover
we assume that the defects like strings and domain wall have the infinite tension being proportional to the
mass of the regulator.  However the naive argument could fail if the very heavy object is dressed 
by the other instanton-like configurations which yields via dimensional transmutation the factor
\beq
\Lambda = M \exp \l -c/g^2(M) \r
\eeq
and potentially could yield nonvanishing contribution.

The issue of instability which would result in expanding is more complicated. It is necessary
to identify the presence or absence of the negative modes at the composite defect which 
is not a simple task. Is there are the odd number of negative modes at the configuration
it would mean that this defect corresponds to the bounce describing the Schwinger-type
process of the creation of the extended object in the graviphoton field.

\subsection{From UV to IR on the defect} 

Recently the interesting approach for the evaluation of the superconformal indexes with the surface
defects of  has been suggested in \cite{rastelli, holland,razamat}. The idea is based
on the realization of the bootstrapping program via particular pattern of  RG flow. The aim is to evaluate
the index  in some quiver-like IR theory with the defect. Instead one enlarges  the theory adding the hypermultiplet
in the bifundamental representation  with respect to the say $SU(N)\times SU(N)$ and consider the 
UV theory first. Since the initial IR theory
is conformal the addition hyper brings the Landau pole into the problem. The FI  term is added to the Lagrangian which forces
the hypermultiplet to condense and fixes the scale in the model.

At the next step one would like to decouple  additional flavor in bifundamental at the scales much lower then one fixed by its
condensate. The decoupling could proceed in two different ways. If the background in UV theory is trivial the
decoupling goes smoothly and we return down to the initial IR theory. However one can select more elaborated way
and start with the nontrivial configuration at UV scale. In \cite{rastelli} one selects the nonabelian vortex configuration 
(see \cite{sy,tong,sakai} for the review). The corresponding condensate of bifundamental becomes inhomogeneous and at IR
the theory becomes the same IR theory without additional hyper but  with additional surface operator which now is the nonabelian string with the infinite tension.

This general picture of RG flows with the nonperturbative defects  turns out to be very useful and provides new tool for the evaluation of indexes. It turns
out that the index of the UV theory allows the integral representation which has the interesting pole structure. 
The residues of the particular poles in the index can be identified with the indexes of IR theory supplemented by the surface operators with some flux $r$. 
There are poles corresponding to the surface operators with the different fluxes. Moreover the index  in the IR theory with the defect with flux r can be identified with the action of the particular difference operator $G_r$ 
with respect to the flavor fugacities acting on the IR index without the defects
\beq
I_r\propto G_r I_0
\eeq
This operator was identified with the Ruijsenaars-Schneider(RS) operator known to be integrable. The trigonometric RS model is nothing but the  CS theory perturbed by two Wilson loops in different directions \cite{gn95}(see Appendix D).

Upon deriving of  the superconformal index in IR theory with defect one could be interested in the additional algebraic structure behind.
It was found in \cite{holland} that the operators $G_r$ form a nontrivial algebra. The surface operators are realized by D2 branes with $R^2\times S^1$ worldsheet and it was demonstrated that the Wilson loop along this $S^1$ emerges 
in the CS theory on $S^1\times C$ where C is the curve defining the superconformal theory. The following correspondence
takes place 
\beq
G_r\leftrightarrow <W_r>
\eeq
where the Wilson loop in the representation r is evaluated. Algebra of the operators $G_r$ gets mapped into 
the Verlinde algebra in the CS theory.

\subsection{Analogy with QCD and CP(N) model}

Let us comment on the related questions which can be raised in QCD and its two-dimensional 
"counterpart" which shares many features of QCD \cite{nsvz1} --  $CP(N)$ sigma-model. Now we know 
well the origin of this correspondence -- it is just the matching condition between the theory in the bulk
and  the worldsheet theory on the defect.
The analogous problem in QCD would concern the Casher-Banks relation for the chiral condensate  relating it 
 with the spectral density of the Dirac operator $\rho(0)$. 
\beq
\ll \bar{\psi}\psi \rr = - \pi \rho(o)
\eeq
The fermionic zero mode at the individual instanton
is senseless in the QCD vacuum since we have strongly interacting instanton ensemble however the collective
effect from the instanton ensemble yields the nonvanishing density at the origin.

Now the question parallel our study would concern the response of the quark condensate on the
"mass of the regulator M"  since we are looking at the derivative of the condensate with respect to 
the mass of the heavy flavor M. 
Speaking differently we are trying to evaluate the effect of the point-like instantons at the Dirac 
operator spectral density. The possible arguments could look as follows. We can consider the well-known
path integral representation for the quark condensate in terms of the Wilson loops 
(see, for instance \cite{migdal})
\beq
 \ll \bar{\psi}\psi \rr= \sum_{paths \quad  C}[DC]  \ll W(C) \rr
\eeq
where the measure over the paths is fixed by the QCD path integral and the vev
of the Wilson loop involves the averaging over all configurations of the gauge field. 
Therefore using this representation we could say that we are searching for the 
back reaction of the UV degrees of freedom at the vev of the Wilson loop. 

Since from our analysis we know that the key players are the point-like instantons
we could wonder how they could affect the Wilson loop.  The natural conjecture 
sounds as follows.
It is known that the Wilson loop renormalization
involves the specific UV contribution from the cusps \cite{polyakov}. Therefore one could imagine that the point-like
instantons placed at the Wilson loop induces the cusps or self-intersections of the Wilson loops 
and therefore yield the additional UV renormalization of the quark condensate.  If this 
interpretation is correct it would imply that the cusp anomalous dimension which on the
other hand carries the information about the anomalous dimensions of the QCD 
operators with the large Lorentz spin should be related with the torus knot invariants.

Another way to approach the question is to use the low-energy theorems \cite{shifman}. The correlator
we are looking at in the SUSY theory is now the correlator of the bilinears of the massless and the
regulator fields. Due to the low-energy theorems we get
\beq
\int d^4x \ll \bar{\psi}\psi(0)\bar{\psi}_R\psi_R(x) \rr \propto  \ll \bar{\psi}\psi \rr
\eeq
which however knows about  the  "perturbative" dilatational Ward identity 
and one could be interested how the point-like instantons affect this relation. We shall see later that the 
similar result
in the SUSY case can be reformulated in the Liouville AGT side  in terms of the similar low-energy theorem "dressed" by the
small instantons.

How similar problem could be posed in the non-SUSY $CP(N)$ model which can appear
as the theory on the defect \cite{gsy}?  The analogue of the closed domain wall considered
above is the kink-antikink bound state which is  true excitation at the large N \cite{witten79}.
The analogous picture looks as follows. We have one vacuum in the non-SUSY $CP(N)$ model however
there is the excited vacuum between the kink-antikink pair. Following our analysis we could conjecture
that this excited vacuum is the analogue of the "regulator vacuum" in the SUSY case separated by the
kinks. Naively it involves the large scale and can just decouple but the kink and antikink can be dressed
by the point-like instantons similar to the dressing of the domain walls by the instantons. As a result
of dressing the  finite $\Lambda$ scale emerges and the kink-antikink state remains in the spectrum.

 Finally note that the chiral 
condensate gets generated in the QED in the external magnetic field \cite{miransky}. Naively it can be traced
from the summation over the lowest Landau level. The analogous question sounds as follows: is there the  interplay
between the fermion condensate and the high Landau levels which are a kind of regulators in this problem. 
Apparently  more involved analysis demonstrated that the
higher Landau levels matter for the condensation  and there is an interplay between the IR and UV physics once again.

\section{Different limits}
\label{sec:lim}
\subsection{Down to HOMFLY, Jones and Alexander}

In order to compare the superpolynomial with other knot invariants, let us rewrite formulas from the previous section in a bit different notation.
It is convenient to change to the following variables:
\begin{eqnarray}
\cfrac{1}{\tq^2 \tt^2} = q = \exp(-\beta \ep_1) \\
\tq^2 = t = \exp(-\beta \ep_2) \\
\ta^2 \tt = A = -\exp(\beta m_a)
\end{eqnarray}

And inverse:
\begin{eqnarray}
\tt = -\cfrac{1}{\sqrt{qt}} = -\exp(\beta(\ep_1+\ep_2)/2) \\
\tq = \sqrt{t} = \exp(-\beta \ep_2/2) \\
\ta = \sqrt{-\sqrt{t q} A} = \exp \l \beta \frac{2m_a-\ep_1-\ep_2}{4} \r
\end{eqnarray}

\begin{itemize}

\item $\tt=-1$: The superpolynomial reduces to HOMFLY. On the field theory side we have $\ep_1+\ep_2=0$

\item $\ta=\tq^N$ corresponds to the quantization condition in NS limit when there is no vev of the scalar. The mass
of the antifundamental gets quantized $m_a=\cfrac{(2N-1)\ep_2 - \ep_1}{2}$ and reduction to the bottom raw or Catalans corresponds
to the semiclassical limit in NS quantization. What is more, if we take $\tt=-1$ we obtain Jones polynomial for the fundamental representation
of $sl(N)$

\item $\ta=1$: we obtain Poincare polynomial for Hegaard-Floer homologies and mass of the anti-fundamental multiplet reads as $m_a=-\cfrac{\ep_1+\ep_2}{2}$. 
Further specification  $\tt=-1$ yields
Alexander polynomial. Hence the sum over Alexanders corresponds to the condensate in the case of massless antifundamental.
It has some interesting realization at the CY side summarized in the Appendix A.

\end{itemize}

\subsection{Up to stable limit}
Consider the limit $k \rightarrow \infty$ which yields the $T_{n,\infty}$ torus knot. At the gauge theory side it corresponds
to the dominance of CS in the action. 

The additional unexpected structure emerges in the stable limit \cite{gor}. It turns out that 
the superpolynomial allows two different "bosonic" and "fermionic" representations. Moreover it
has the structure of the character of the very special representation in $\hat{SL(2)}$ at level 1
introduced by Feigin and Stoyanovsky. 

Where  $\hat{SL(2)}$ at level 1 could appear from? The possible conjecture sounds as follows. We 
have to recognize the knot invariants from the viewpoint of the theory on the flavor branes as well.
 two hypermultiplets, one in fundamental and one in antifundamental. The theory on their 
worldvolumes should enjoy $SL(2)$ gauge group instead of SU(2) for two fundamentals.
It is the analogue of the Chiral Lagrangian realized as the worldvolume theory on the flavor branes. Similar to 
the QCD we have the CS term here as well and the coefficient in front of it equals to the number of colors.
In our abelian case  we immediately arrive at the level 1 as expected. The "Chiral Lagrangian" in our case
could have Skyrmion solutions like in QCD and we conjecture that the Feigin-Stoyanovsky representation
is just representation in terms of Skyrmions.

Another question which can be asked in the stable limit is the A-polynomial. The point 
is that superpolynomial of uncolored $T_{n,\infty}$ torus knot gives superpolynomial of unknot 
colored by the n-th symmetric representation $M_n$: 
\beq
P_{n,\infty}(a,q,t) = \cfrac{M_n(a,q,t)}{M_1(a,q,t)}
\eeq
These are given by the so-called MacDonald dimensions\cite{dunin}. Therefore we could investigate
the dependence of the superpolynomial on the representation which is governed
by the super-A-polynomia l\cite{gukov12} of the knot. For the unknot in the symmetric representation, in our normalization it reads as
\beq
\hat A(a,q,t,\hat x,\hat y)=\cfrac{t a \hat x - a^{-1} \hat x^{-1}}{\hat x q- \hat x^{-1} q^{-1}}+\sqrt{-t} \hat y
\eeq
where operators $\hat x, \hat y$ act as
\beq
\hat y M_n = M_{n+1}, \ \hat x M_n = (-tq)^n M_n
\eeq
and quantum A-polynomial annihilates superpolynomial:
\beq
\hat A M_n = 0
\eeq
Note that this equation actually connects two different instanton contributions $M_n$ and $M_{n+1}$. Recall that in the NS limit $\ep_2=0$ but for a generic $k$, we have found similar relation
(\ref{airy}) which relates different instanton contributions too. These relations are similar in the spirit to non-perturbative Dyson-Schwinger equations introduced recently 
by N. Nekraso v\cite{nekrasov14}.
We hope to discuss this issue elsewhere \cite{bgmn}.

\subsection{ Nabla - Shift  - Cut-and-join operator and the decoupling of heavy flavor}

Let us make  few comments concerning the role of  operator providing the transformation
$k \rightarrow k+1$ in our picture. It has different reincarnations and different names  in the several 
problems. It is known as Nabla operator in the theory of the symmetric functions, as the
shift operator in the rational DAHA and Calogero model and cut-in-join operator
in the context of some counting  problems in geometry. Since we have identified this
parameter as the level of 5D CS term we could use this knowledge and see the different
interpretations of this shift.

From the physical side it can be immediately recognize as the effect of the decoupling 
of the heavy flavor since it is know for a while \cite{witten96} that one-loop effect provides
this shift of the level. It provides the shift of the Calogero coupling constant in the quantum
mechanics on the instanton moduli space and more formally it corresponds to the 
multiplication of the integrand over the instanton moduli by the determinant bundle \cite{tachikawa}.
It can be also seen at the CY side when it corresponds to the change of the geometry.

In the consider action on the $T_{n,nk+1}$ torus knots this operator was used  \cite{dunin} to
generate  a kind of the discrete Hamiltonian evolution in k  with the simple boundary condition for k=0,
corresponding to unknot. Having in mind that the dynamics of the Calogero coupling can be interpreted
as the realization of the RG evolution with the limit cycles \cite{gorskyrg} it would be interesting to look for the
cyclic solutions to this discrete Hamiltonian dynamics and possible Efimov-like states in this
framework.

To illustrate these arguments  let us consider the limit $m_a \ra \infty$, that is $A \ra +\infty$. From physical
viewpoint, we integrate out antifundamental multiplet, so the Chern-Simons coupling should reduce by one. Indeed, the following limit is well-defined:
\beq
\lim_{A \ra +\infty} \cfrac{1}{A^{n-1}} P(A,q,t)_{nm+1,n} = P(A=0,q,t)_{n(m-1)+1,n}
\eeq
Actually, this relation is known in the knot theory and is quite general:
\beq
\lim_{A \ra +\infty} \cfrac{1}{A^{n-1}} P(A,q,t)_{k,n} = const P(A=0,q,t)_{k-n,n}
\eeq

The cut-and-join operator was identified in \cite{dunin} as the $W^3_0$  generator from $W^3$ algebra.
This fits well with the our consideration since the CS  term written in superfield looks as follows 
\beq
\delta L_{CS}= \int d^5x d^4 \theta \Phi^3
\eeq
It is possible to develop the matrix model of the Dijkgraaf-Vafa type for the 5D gauge theory  \cite{klemm08} which can be 
considered as the generation function for the  superpolynomials of the $T_{n,nK+1}$ torus knots. 
The matrix model evaluation of the particular observable in the general case of all
nonvanishing masses looks as 
\beq
Z_{matr}= \int [d M] O(m_1)O(m_2)O(m_3) exp (t_2TrM^2 +t_3 TrM^3)
\eeq
with some measure probably suggested in \cite{agash} and the coefficient $t_3$ corresponds 
to the CS term. It is clear that in the matrix model framework it is coupled to the corresponding
$W^3_0$  generator. The knot invariants presumably can be evaluated upon taking derivatives with
respect to $m_1, m_2$ and the corresponding limits. The operators O(m) are conjectured to be 
\beq
O(m) = det(M-m) 
\eeq
which can be evidently related with the resolvents. The
type of the knot presumably is selected by the corresponding term of expansion in $t_2$
with fixed value  of $t_3$.

Therefore the  shift  operator  can be thought of as one of the consequences from the
generalized Konishi relation in the 5D gauge theory yielding the W-constraints in the
matrix model. If one introduces more times more general Virasoro and W- constraints
can be formulated for the torus knots superpolynomials (see \cite{dubinkin} for the
related discussion). Let us emphasize that 
the matrix model with the cubic potential is different from the matrix model developed
for the evaluation of the torus knot invariants from the type B topological strings
in \cite{einard,klemm}.

\section{AGT conjecture prospective}
\label{sec:agt}
In this section we will continue our study of the five-dimensional SQED with two fundamental flavors in the $\Omega$ deformation. 
We will show in a moment that the \textit{instanton} partition function for 5D SQED is directly related to the \textit{perturbative}
partition function for 5D $SU(2)$ gauge theory. Therefore, according to the AGT conjecture\cite{agt}, there should be a
relation between three-point functions in Liouville theory and five-dimensional SQED. 
We will establish an explicit
connection between 5D SQED and the three-point function in the q-deformed Liouville theory on a sphere.
We argue that the  three-point function is equal to the combination of four instanton partition functions. Using these results we
will show that there is a relation between superpolynomials for unknot and Liouville theree-point functions.

\begin{figure}[!h]
\begin{center}
\setlength{\unitlength}{1.2cm}
\begin{picture}(7,6)
\linethickness{0.3mm}
\put(2,2){\line(1,0){2}}
\put(4,2){\line(0,1){1}}
\put(4,2){\line(1,-1){1}}
\put(5,1){\line(1,0){1}}
\put(5,1){\line(0,-1){1}}
\put(4,3){\line(0,1){1}}
\put(2,4){\line(1,0){2}}
\put(4,4){\line(1,1){1}}
\put(5,5){\line(0,1){1}}
\put(5,5){\line(1,0){1}}
\put(2,2){\line(0,1){2}}
\put(2,2){\line(-1,-1){1}}
\put(2,4){\line(-1,1){1}}
\put(1,1){\line(-1,0){1}}
\put(1,1){\line(0,-1){1}}
\put(1,5){\line(-1,0){1}}
\put(1,5){\line(0,1){1}}
\put(2.9,1.7){\makebox{$Q$}}
\put(4.7,4.4){\makebox{$Q_{m_3}$}}
\put(4.1,3.0){\makebox{$Q_c$}}
\put(4.5,1.5){\makebox{$Q_{m_2}$}}
\put(1.6,1.2){\makebox{$Q_{m_1}$}}
\put(0.7,4.2){\makebox{$Q_{m_4}$}}
\end{picture}
\end{center}
\caption{$SU(2)$ theory with four flavours}
\label{fg:su2}
\end{figure}
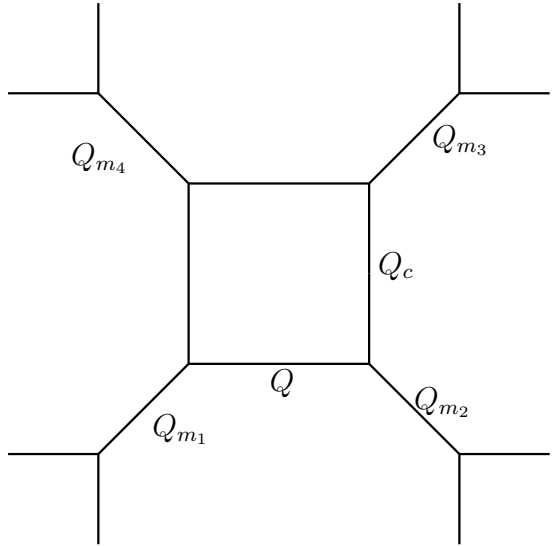

Let us consider $D=5, \ \mathcal{N}=1 \ SU(2)$ gauge theory with four flavours. It can be obtained as a theory living 
on a certain IIB $(p,q)$-brane web - see Figure \ref{fg:su2}.  Or, equivalently, as M-theory compactification on the toric Calabi-Yau theefold,
those toric diagram is given by the same figure.

\begin{figure}[!h]
\begin{center}
\setlength{\unitlength}{1.2cm}
\begin{picture}(7,6)
\linethickness{0.3mm}
\put(2,2){\line(1,0){0.5}}
\put(3.5,2){\line(1,0){0.5}}

\put(4,2){\line(0,1){1}}
\put(4,2){\line(1,-1){1}}
\put(5,1){\line(1,0){1}}
\put(5,1){\line(0,-1){1}}
\put(4,3){\line(0,1){1}}
\put(2,4){\line(1,0){0.5}}
\put(3.5,4){\line(1,0){0.5}}
\put(4,4){\line(1,1){1}}
\put(5,5){\line(0,1){1}}
\put(5,5){\line(1,0){1}}
\put(2,2){\line(0,1){2}}
\put(2,2){\line(-1,-1){1}}
\put(2,4){\line(-1,1){1}}
\put(1,1){\line(-1,0){1}}
\put(1,1){\line(0,-1){1}}
\put(1,5){\line(-1,0){1}}
\put(1,5){\line(0,1){1}}
\put(4.7,4.4){\makebox{$Q_{m_3}$}}
\put(4.1,3.0){\makebox{$Q_c$}}
\put(4.5,1.5){\makebox{$Q_{m_2}$}}
\put(1.6,1.2){\makebox{$Q_{m_1}$}}
\put(0.7,4.2){\makebox{$Q_{m_4}$}}
\end{picture}
\end{center}
\caption{Perturbative $SU(2)$ theory with four flavours}
\label{fg:pert}
\end{figure}
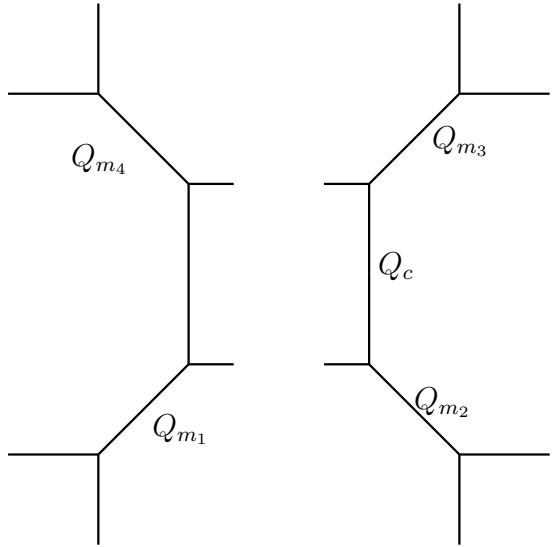

We can obtain two copies of 5D SQED by sending Coloumb parameter to infinity $a \ra +\infty(Q_c \ra 0)$, that is, by cutting
the toric diagram by horizontal line - compare\footnote{Note that we obtain two $U(1)$ theories with CS term $\pm 1$. This is not surprising since we have to integrate out
heavy chiral gaugino} the result with Figure \ref{fg:k1}.
If we send $g^2 \ra \infty(Q \ra 0)$ then the partition function is given only by the perturbative contribution - we cut the toric
diagram by vertical line - see Figure \ref{fg:pert}

Note that the $SU(2)$ theory has a rotational symmetry(fiber-base duality)\cite{fb1,fb2,bao11} which interchanges 
coupling constant $Q=\exp(-\beta/g^2)$ with 
Coloumb parameter $Q_f=\exp(-\beta a)$. It is this symmetry which relates perturbative $SU(2)$ theory with instanton $U(1)$ - one has
to rotate Figure \ref{fg:pert} by 90 degrees in order to obtain Figure \ref{fg:k1}

In the pioneer work \cite{agt} it was shown that the perturbative part of the Nekrasov partition function for the four-dimensional 
$SU(2)$ theory with four fundamental flavors actually coincides with the three-point function 
$\ll e^{2 \al_1 \phi} e^{2 \al_2 \phi} e^{2 \al_3 \phi} \rr$ in the Liouville theory on a sphere, also known as a
DOZZ factor \cite{dozz1,dozz2}:
\begin{equation}
\begin{split}
 C(\al_1,\al_2,\al_3)=(\pi \mu \gamma(b^2) b^{2-2b^2})^{(Q-\al_1-\al_2-\al_3)/b} \times \\
\cfrac{\up'(0) \up(2\al_1) \up(2\al_2) \up(2\al_3)}{\up(\al_1+\al_2+\al_3-Q_c) \up(\al_1+\al_2-\al_3) \up(\al_1+\al_3-\al_2) \up(\al_2+\al_3-\al_1)}
\end{split}
\end{equation}
where we have adopted the standard notation for the Liouville theory: the central charge is given by $c=1+6Q_c^2$, $Q_c=b+1/b$. And $\up$ is a combination
of two Barnes' functions:
\begin{equation}
 \up(x)= \cfrac{1}{\Gamma_2(x|b,\bm) \Gamma_2(Q-x|b,\bm)}
\end{equation}
One can think about the Barnes' function $\Gamma_2$ as a regularized product:
\begin{equation}
\Gamma_2(x|\ep_1,\ep_2)=\prod_{n,m=0}^{+\infty}(x+m \ep_1+ n \ep_2)^{-1}
\end{equation}
 - for the case of the combination $\cfrac{\Gamma_2(x) \Gamma_2(y)}{\Gamma_2(x+z) \Gamma_2(y-z)}$ this is a precise prescription.

In \cite{awata} and later in \cite{aandb,wyllard,bao11} it was argued that the lift to the five-dimensional theory corresponds to the q-deformation
on the CFT side. Now we are going to extend the proposal of \cite{bao11}, which connects the full(including both perturbative and
non-perturbative) Nekrasov partition function for the 5D Abelian theory with two flavors with the q-deformed DOZZ factor in the case $c=1$. We argue
that the same relation holds for the general central charge. However, in our approach we will need the combination of several instanton partition
functions in order to obtain a single DOZZ factor. 

First of all, let us recall the q-deformation of Barnes' double gamma function, which is closely related to the MacMahon function.
Again, we will not need a precise definition, since we are interested in rations of four such
functions:
\beq
\Gamma^\beta_2(x|\ep_1,\ep_2)= \prod_{i,j=0}(1-\exp(-\beta(x+i \ep_1 + j \ep_2)))^{-1}
\eeq
And correspondingly:
\beq
 \up^\beta(x)= \cfrac{1}{\Gamma^\beta_2(x|b,\bm) \Gamma^\beta_2(Q-x|b,\bm)}
\eeq
Following \cite{bao11}, we define q-deformed DOZZ factor by substituting Barnes' functions by their q-analogues. However, we will omit the factor 
$(\pi \mu \gamma(b^2) b^{2-2b^2})^{(Q-\al_1-\al_2-\al_3)/b}$ since it can be absorbed into the definition of vertex operators:
\beq
C^\beta(\al_1,\al_2,\al_3)=
\cfrac{\up^{\beta'} (0) \up^\beta (2\al_1) \up^\beta (2\al_2) \up^\beta (2\al_3) }
{\up^\beta (\al_1+\al_2+\al_3-Q_c) \up^\beta (\al_1+\al_2-\al_3) \up^\beta (\al_1+\al_3-\al_2) \up^\beta (\al_2+\al_3-\al_1)}
\eeq

Returning to the 5D partition function, 
it will be useful to consider a bit different representation for the partition function from the section \ref{sec:qt}\cite{iqbal08}:
\beq
Z_{inst}=\prod_{i,j=0} \cfrac{(1-\frac{Q Q_a}{\sqrt{qt}} q^{i+1/2} t^{j+1/2})(1-Q Q_f \sqt q^{i+1/2} t^{j+1/2})}{(1-Q q^{i+1/2} t^{j+1/2})(1-Q Q_f Q_a q^{i+1/2} t^{j+1/2})}
\eeq
where we have used an analytic continuation:
\beq
\prod_{i,j=1} (1-Q q^{i-1/2} t^{j-1/2}) = \prod_{i,j=1} (1-Q q^{i-1/2} t^{1/2-j})^{-1}
\eeq
K\"ahler parameters for the coupling constant, fundamental and antifundamental masses read as:
\beq
Q=\exp(-\beta/g),\ Q_f = \exp(-\beta m_f)/\sqt = \mu_f/\sqt, \ Q_a = \sqt \exp(\beta m_a) = \sqt \mu_a
\eeq
Finally, the partition function can be rewritten as:
\beq
Z_{inst}(1/g,m_f,m_a)=\cfrac{\Gamma^\beta_2(1/g - m_a + \ep_1/2 + \ep_2/2|\ep_1,\ep_2) \Gamma^\beta_2(1/g+m_f + \ep_1/2 + \ep_2/2|\ep_1,\ep_2)}
{\Gamma^\beta_2(1/g  + \ep_1/2 + \ep_2/2|\ep_1,\ep_2) \Gamma^\beta_2(1/g + m_f - m_a  + \ep_1/2 + \ep_2/2|\ep_1,\ep_2)}
\eeq

We see that the DOZZ factor and the 5D partition function are strikingly similar. First of all, we can establish usual relation in AGT correspondence:
\beq
b = \ep_1, \bm = \ep_2, Q_c = \ep_1 + \ep_2
\eeq
Then, it is straightforward to obtain the following expression:
\beq
\up^{\beta'} (0)  = \cfrac{1}{\beta} \prod_{i,j=0} (1-q^i t^j)_{0,0} (1-\exp(-\beta Q_c) q^i t^j)
\eeq
where the subscript $0,0$ denotes the omission of the $i=j=0$ term. 

After trivial manipulations with various factors we arrive at the following identification:
\beq
\label{dozz_wave}
C^\beta(\al_1,\al_2,\al_3)=\cfrac{1}{Q \sqt}\cfrac{Z^3_{inst} \cfrac{\pr Z^1_{inst}}{\pr m_f}}{Z^2_{inst} Z^4_{inst}}
\eeq
where
\begin{eqnarray}
Z^1_{inst}=Z_{inst}(\gm=\hqc+\al_1-\al_2-\al_3,m_f=-\gm-\hqc,m_a=\gm+\hqc-2 \al_1) \\
Z^2_{inst}=Z_{inst}(\hqc-2 \al_1, \al_2+\al_3-\al_1-\gm-\hqc,\gm+\al_1+\al_2+\al_3-3\hqc)\\
Z^3_{inst}=Z_{inst}(\al_1+\al_2-\al_3-\hqc,2 \al_2-\gm-\hqc,\gm+2 \al_3-\hqc) \\
Z^4_{inst}=Z_{inst}(2\al_3-\hqc,\al_3+\al_1-\al_2-\gm-\hqc,\gm+\al_2+\al_1-\al_3-\hqc)
\end{eqnarray}

Equation (\ref{dozz_wave}) suggests that the DOZZ function is equal to the composite defect wave function, since the derivative with respect 
to $m_f$ corresponds to the insertion of this defect, whereas the wave function is literary equal to the partition function in the presence of the defect. Terms in the denominator are conjugate wave 
functions.

Now let us return to the torus knot superpolynomial. It is clear that the derivative with respect to the fundamental mass
corresponds to correlators of the form $\ll \phi e^{2 \al_1 \phi} e^{2 \al_2 \phi} e^{2 \al_3 \phi} \rr$ on the Liouville side. However, the role
of the operator $\exp(-\beta \Phi)$ is not quite clear. We will show now that in the absence of the CS term($k=0$), it is not necessary 
to consider the VEV of $\exp(-\beta \Phi)$, since this VEV and
the partition function is actually proportional. This observation establishes a bridge between Liouville correlators and torus knots.

Let us consider the following peculiar observable\cite{nekrasov14}:
\beq
\label{bethe}
\ll Y(qtz) + Q \cfrac{z^k(z-\exp(\beta m_{af}))(1-z \exp(-\beta m_{f})) }{Y(z)} \rr
\eeq
where $Y(z)$ is a generating function for chiral ring observables:

\beq
Y(z)=(z-1)\exp \l \sum_{n=1}^\infty \cfrac{z^{-n} \ll \exp \l -\beta n \Phi \r \rr}{n} \r
\eeq

Also, in \cite{ns092,strings09} 
it was conjectured that the operator $Y(z)$ corresponds to the insertion of a domain wall. For a particular instanton configuration,
defined by a Young diagram $\la$, $Y(z)$ equals to
\beq
Y(z) = \cfrac{\prod_{\pr_+ \la}(z-q^{a'} t^{l'})}{\prod_{\pr_- \la}(z-qt q^{a'} t^{l'})} = z\cfrac{\prod_{\pr_+ \la}(1-q^{a'} t^{l'}/z)}{\prod_{\pr_- \la}(1-qt q^{a'} t^{l'}/z)} 	
\eeq
where $\pr_+ \la$ defines cells which can be added to the Young diagram and $\pr_- \la$ are those which we can remove.

According to N. Nekrasov\cite{nekrasov14},  (\ref{bethe}) is a regular function as a function of "spectral parameter" z. This is an analogue of Baxter TQ-equation for 
general Omega-deformation. In our case it is a polynomial of degree $k+1$.
If we use the identities
\begin{eqnarray}
1-(1-q)(1-t)\sum_\Box q^{a'} t^{l'} = \sum_{\pr_+ \la} q^{a'} t^{l'} - qt \sum_{\pr_- \la} q^{a'} t^{l'} \\ \nonumber
Y(0)=-1
\end{eqnarray}
Then we see that
\beq
\ll Y(z) \rr = z Z_{inst} - Z_{inst} + \ll \exp(-\beta \Phi) \rr + O(1/z),\ z \ra \infty
\eeq
Let us concentrate on the case $k=0$ which corresponds to the unknot.
We can find the constant term in (\ref{bethe}) by taking $z=0$:
\beq
-Z-Q Z \exp(\beta m_{af})
\eeq
On the other hand we obtain this term by taking $z \ra \infty$. Finally, we arrive at:
\beq
\label{exp_z}
\ll \exp(-\beta \Phi) \rr = Q Z \cfrac{\emf - \emaf - \emaf \emf -1 }{1+Q \emf}
\eeq
Recalling that the instanton partition function $Z=\ll 1 \rr$ is 
equal to 1 if $m_f$ or $m_{af}$ equal to zero, we obtain:
\beq
\label{pol_z}
 -\cfrac{\pr}{\pr( \beta  m_f)} \ll \exp(-\beta \Phi) \rr {\Big \vert}_{m_f=0} = Q \cfrac{\pr Z}{\pr( \beta m_f)} \cfrac{2 \emaf}{1+Q} - Q \cfrac{\emaf(1-Q)-(1+Q)}{(1+Q)^2}
\eeq
The problem with $k=1$ is that one has to consider higher-order terms in the expansion of $Y(z)$.

\section{ Knot invariants from  quantum mechanics on the instanton moduli space  and $n \leftrightarrow m$ duality}
\label{sec:qm}
There are complicated consistency conditions for the branes of  different dimensions  to live together happily and 
they are formulated differently in their worldvolume theories.
All physical phenomena have to equivalently described from the viewpoints of the worldvolume theories
on the defects of the different dimensions involved.  Therefore we have to recognize the knot invariants in the  corresponding  quantum mechanics on the instanton moduli space.
In this Section we consider the NS limit postponing the case of the general $\Omega$- background for the further study. As we have mentioned before one could expect that the
states near threshold should matter for the UV completion problem and we shall see that it is indeed the case.

Let us remind, following \cite{gordaha} how the Poincare polynomial of the HOMFLY homologies of the torus knots are obtained in the Calogero model.
To this aim it is useful to represent the quantum Calogero Hamiltonian in terms of the Dunkl operators
\beq
H_{cal}= \sum_n\partial _n^2 + \sum_{i\neq n}\frac{c(c-1)}{(z_i -z_n)^2}
\eeq
\beq
H_{cal}=\sum_n D_n^2 \qquad D_n= \partial_n + c\sum_{i\neq n}\frac{1-\sigma_{i,n}}{z_i -z_n}
\eeq
The Dunkl operators enters as generators in the rational DAHA algebra  \cite{cherednik}. It is important
that for rational Calogero coupling $c=n/m$ there is the finite-dimensional representation of DAHA \cite{berest}. It is this finite-dimensional representation does the job. 

It was shown in \cite{gors} that the particular twisted character of this finite-dimensional representation coincides with
the Poincare polynomial of the HOMFLY  homology of the $T_{n,m}$ torus knot
\beq
P_{n,m}(a,q)= a^{(n-1)(m-1)}\sum_{i=0}^{n-1}a^{2i}tr(q^{\rho};Hom_{S_n}(\Lambda^i h, L_{m/n}))
\eeq
where the following objects are involved. The $h$ is the $(n-1)$ dimensional reflection representation of $S_n$, $C[h],C[h^{*}],C[S_n]$ generates the whole Cherednik algebra (we present its definition in Appendix).
The $L_{m/n}$ is the finite-dimensional representation of the Cherednik algebra. The $n\leftrightarrow m$ symmetry
is not evident however it was proved in \cite{gordaha} via comparison with the arc spaces on the Seifert surfaces
of the torus knots. The arc space is  nothing but  the  space of open topological string instantons in the physical language.  
The element
$\rho$ belongs to the algebra and acts semisimply. Its eigenvalues provides the q-grading in the character representation.
It can be thought of as the Cartan element of $SL(2,R)$ subalgebra of the Cherednik algebra which is known for the Calogero
model and plays the role of the spectrum generating algebra.
This Cartan element corresponds to  one of the U(1) rotations of the $\mathbb{C}^2$ where the instantons live.

Therefore from the Calogero viewpoint the HOMFLY invariant can be considered as a kind of generalization
of the Witten index.  The HOMFLY torus knot invariants are captured by the subspace of the 
rational complexified Calogero model  Hilbert space.
As we argued before the torus knot invariants are derived upon the integration of the determinants over the instanton moduli space in 5d gauge theory and
the integral is localized at the  centered instantons at one point. This fits with the relevance of the case when all
Calogero particles are concentrated around the origin and we consider a kind of the "falling at the center" problem.  Note
that the rational Calogero system is the conformal quantum mechanical  model and we effectively impose the restriction on the spectrum.

Where the  Calogero model  with the particular coupling comes from in our instanton problem? The answer comes from the 
quantum mechanics on $n$-instantons moduli space. The small abelian instantons yield the Calogero model indeed if 
we think about  the theory on the commutative space when some number of the points are blow-uped \cite{braden}. If the 
abelian instantons are restricted on the complex line one gets the Calogero model for the elongated
instantons indeed as shown in \cite{braden}. 

 We have to explain why the coupling constant in the Calogero model equals to $n/m$.
The key point is that the CS term induces the magnetic field on the ADHM  moduli space  \cite{collie,kim} which is equal to the
level of CS term, in our case $B_{eff}= k$.
Immediately we can recognize that the coupling constant in the Calogero model
corresponding to the $T_{n,nk+1}$ is the CS level $k=\frac{nk+1}{n}$ at least at large $n$  as required from the DAHA representation. Therefore we could claim that it is CS term which
generates the correct  interaction of Calogero particles. 

The proper framework to explain $n \leftrightarrow m$ duality in the Calogero coupling is suggested by the QHE which can be 
approximately described by the $n$-body  Calogero or RS models depending on geometry \cite{susskind,poly,gkk}. 
In the Calogero approach to QHE the Calogero coupling is equal to the filling factor which is related to the 
coefficient in front of the 3d CS term in the effective theory of the integer QHE. This is parallel to our case where the Calogero interaction is 
induced via the reduction of 5d CS term to 3d and the Calogero coupling is the level of CS term again. Fermions in the IQHE get substituted by the abelian instantons in our case.
With this identification of the Calogero model we could expect that duality in the torus knot problem  gets 
mapped into the similar duality in the integer QHE. The $\nu \rightarrow \nu^{-1}$ duality in IQHE corresponds to the substitution of  quasiparticles by  holes and vise versa.

It is in order to make some digression which can be interesting by its own. In our study we have started with the theory
with $N_f=3$ which has Landau pole. The theory with $N_f=2$ has vanishing $\beta$- function while theory with $N_f=1$ is asymptotically free. 
They are very different therefore we could look for the origin of this difference in the our framework.
We know that one flavor is massless therefore we can not decouple it however the limit $A \rightarrow 0$ provides the
decoupling of one flavor. The transition from $N_f=3 \rightarrow N_f=2$   from the  point-like instantons looks quite smooth.
Starting with all massive flavors   and sending  one mass to infinity we see the clear picture of the knotting 
and formation of some small compact UV  defect where the small instantons are nested. That is transition 
from the theory with Landau pole to the conformal theory involves the formation of some small size UV defect.

We can look at this point from the slightly different angle, namely from the realization of the HOMFLY polynomial in terms of the finite-dimensional 
representation of rational DAHA. When $A \neq 0$ we have conformal regime perturbed by the
particular nonlocal operator. In this case the parameter a which we have identified with the mass of the fundamental
measures the representation of $S^n$ on the $n$-point-like instantons- Calogero particles. Since mass is related to the Casimir
of the translation $(p^2=m^2)$ we could claim that the symmetric group and the Lorentz group are related in the nontrivial
way. Indeed the $SL(2,R)$ rotations are built in the rational DAHA nontrivially. 

However the momentum is not the generator of the DAHA therefore the mixing of the $n$-particle state under the translations
occurs and the different representations of the symmetric group emerge. When $A \rightarrow 0$  a bit unusual counting of the 
representation of the symmetric group by mass serving as fugacity disappears that is in some sense the group of space-time translations and 
symmetric group decouple from each other. This seem to be some important property of the asymptotically free theories which has to  be elaborated in more details.

\section{Conclusion}

In this paper we have formulated the explicit instanton--torus knot duality between the $n$-instanton contribution to the
particular condensate in the 5D SQED and superpolynomial of the particular torus knot. The second derivative 
of the Nekrasov partition function plays the role of the generating function for the torus knot superpolynomials.
The condensate is evaluated in the background of the 4-observable - which can be considered  as the result of the  incomplete decoupling of the regulator  degree of freedom. Hence to some extend we could say that the knot invariants govern the delicate UV properties of the gauge theory when the point-like instantons interact with the UV degrees of freedom.

What are the lessons  we could learn from this  correspondence? Some of them have  been already mentioned in the Introduction.
First of all 
the appearance of the higher $q,t$-Catalan number tells that we dealing with the point-like instantons sitting at the top of each other instead of the randomly 
distributed in $\mathbb{R}^4$. This effect is due to the regulator degrees of freedom which
yields the non-local operator in the correlator. Secondly the non-locality of the operator induced by the UV regulator degree
of freedom  implies that we have to recognize the compact nonlocal object. 
This is the candidate state for the "elementary rotator" and on the other hand it 
captures the information on the torus knots superpolynomials.  

The other lesson can be formulated as follows. The example in \cite{rastelli} demonstrates
that the UV degrees of freedom can penetrate as the non-Abelian strings with the infinite tension known as the
surface operators. Similar logic can be applied for the domain walls  which separates the region with the "UV" degrees
of freedom and region with the IR degrees of freedom only. In the similar simplest case
the  domain wall tension becomes infinite when the mass of the regulator tends to be infinite and the situation 
is analogous to the string with the infinite tension. Such domain walls yields the boundary conditions only. 

However one could consider the more interesting question: are there the compact defects involving the UV regulator 
degrees of freedom with the finite action. The candidates are the closed string and closed domain wall. In this 
case we have to find arguments preventing them from shrinking to a point and therefore escaping from the
physical spectrum. If considering the closed string involving the regulators the first mechanism  could be analogue
of the mechanism considered by Shifman and Yung in the "instead-of-confinement" approach  \cite{instead}when the quantum 
state of the monopole-antimonopole pair nested on the string prevent it from shrinking. This monopole-antimonopole pair
transforms into the interesting state upon the Seiberg duality.  Another way to stabilize the closed string is to 
add a kind of rotation  due to the additional quantum number like for the Hopf string considered in \cite{gsy13}.
Similar mechanisms can be applied to the closed domain wall as well. One can consider its stabilization   
via the defects of low dimension or by rotation induced by the additional quantum numbers. There is for instance the "monopole bag" configuration \cite{bolognesi}. 
In our case we have a kind of such object which seems to be prevented from shrinking by the instantons
inside. Moreover one could expect that the   mass of the regulator which enters the domain wall tension is "dressed" by the point-like instantons and its transforms into the $\Lambda_{QCD}$ -type scale via dimensional transmutation thus 
providing the some dynamical mechanism behind it.

One more  important  physical lesson  to be  learned is as follows. In the conventional QCD there is the fermionic zero mode at the individual instanton, however 
the fermionic chiral condensate is not due to it. The chiral condensate is determined via Casher-Banks relation
by the density of the quasi-zero modes in the instanton-anti-instanton ensemble. The details of the interaction of the
instantons can not be recognized in this case and only the collective effect can be seen in Casher-Banks relation. 
In our case due to supersymmetry we can say more on the microscopic structure behind the condensate. The CS  term induces the attractive interaction of Calogero-type between 
point-like instantons  and the falling at the center phenomena occurs. It turns out that the accurate treatment of this phenomena is performed  in terms of the knot invariants. In our case the torus 
knots are selected by the choice of the matter content, however  in general more complicated knots can be expected.
The summation over the torus knot superpolynomials amounts to the  nonvanishing  condensate and corresponds to the 
summation over instantons which to some extend yields the microscopic picture for the analogue of  "Casher-Banks" relation.

It seems that we just have touched the tip of the iceberg and  there are  immediate questions to be formulated.

\begin{itemize}

\item How the matter content of the 5D has to be extended to fit with the general $T_{n,m}$ torus knots and links? We expect  that the superpotential in the 5d theory is in  one-to -one correspondence with the type of the knot. The torus knots correspond to the simplest case when only the CS term is involved. 

\item How the instanton-knot correspondence will be modified in the nonabelian case and for the general quiver theory?

\item We have not discuss in this study the differentials  in the Khovanov homologies postponing this issue for the separate work. They should
be related to the effect of surface operators. Indeed, it was shown in \cite{gors} that the differentials are attributed
to the complex lines in $R^4$. The related question concerns the derivation of the
Hall algebra in the 5d theory. The results in \cite{gs} certainly should be of some use.

\item What is the meaning of the fourth grading which seems to be under the carpet  for the
torus knots  \cite{ggs13} in the 5D gauge theory?

\item What amount of this duality survives in the 4d case?

\item We have not touched in this paper the theory on the worldvolume of the flavor brane at all postponing this issue 
for the future work. This would involve the analogue of the Chiral Lagrangian and we will have to recognize the
torus knot invariants from this perspective as well. It is known that the Skyrmion is represented as the 
instanton trapped by the domain wall in D=5  gauge theory \cite{sakai} realizing dynamically the Atyah-Manton picture.
It seems that  our composite defect could have a relation with a kind of Skyrmion or dyonic Skyrmion
in the "Chiral Lagrangian".

\item It seems that our considerations have common features with the Higher dimensional \cite{4dQHE} QHE
effect in the refined case  and the conventional QHE in the unrefined case. Is it possible to clarify the role of the torus knot
invariants in that context?

\item Is it possible to have the simple interpretation of $n\leftrightarrow (nk+1)$ duality in terms of the
topological strings and in the theory on the surface operator?

\item Recently the relation between the RG cycles and the decoupling of the heavy degrees of freedom
has been found in the $\Omega$-deformed SQCD. The similar RG cycles were found in the Calogero model \cite{gorskyrg,bgrg} in this context.
How these RG cycles can be formulated in terms of knot invariants? Are there the Efimov-like states?

\item There is an interesting duality between the pair of integrable systems. One classical integrable system
belongs to the Toda-Calogero-RS family while the second quantum integrable model is a kind of the spin chain.
The mapping between two sides of the correspondence is quite nontrivial \cite{givental, gk,gzz}. In our case we see that the
knot invariants are related with the spectrum of the quantum Calogero system. Is it possible to recognize
the knot invariants at the spin chain side when the additional deformation is included? Some step in this direction
was made in \cite{bg14}.

\item Recently an additional clarification of the 2d-4d duality has been obtained. Using the representation
of the nonabelian string in terms of the resolvent \cite{ggs} in N=1 SYM theory the issue of the gluino condensate
has been reconsidered in \cite{sy14}. It was shown, using the interplay between 2d and 4d generalized Konishi anomalies, that the gluino condensate in N=1 theory 
penetrates the worldsheet theory and deforms the chiral ring and corresponding
Bethe ansatz equations in the worldsheet theory in the nontrivial manner. Is it possible to use the 5d-3d correspondence
to recognize the knot invariants on the defect "inside the condensate"?

\item How the critical behavior discussed in \cite{bgn14} will be generalized in our situation \cite{bgmn} and what is its
proper physical interpretation?

\item Is it possible to make the arguments concerning the explanations of the  dimensional transmutation 
phenomena via the composite defect precise?

\item How these composite defects interact?

\end{itemize}
We hope to discuss these issues elsewhere.

\section*{Acknowledgment}
We would like to thank  K.~Bulycheva, I.~Danilenko, M.~Gorsky, S.~Gukov, S.~Nechaev, A.~Vainshtein and especially E.~Gorsky and 
N.~Nekrasov for the useful discussions.
The research was carried out at the IITP RAS at the expense of the Russian Foundation for Sciences (project \textnumero \ 14-50-00150).
\section{Appendix A. Torus knots}
\begin{figure}[h]
\begin{center}
\centering
\includegraphics[scale=0.5]{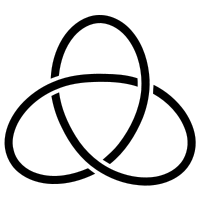}
\caption{Trefoil knot(this figure courtesy of \textit{Wikipedia})}
\label{fg:kn32}
\end{center}
\end{figure}

In this Appendix we will sketch some properties of torus knots. For a comprehensive review, see \cite{kauffman}.
The general definition of a knot is a continuous embedding of $S^1$ into $S^3$ up to a homotopy. The trivial example in unknot:  this is
just a circle lying inside $S^3$. According to the Thurston theorem\cite{thur}, every knot is either:
\begin{itemize}
\item Satellite. Such knots can be obtained by taking a non-trivial knot lying inside a solid 2-torus(in this situation non-trivial means that the knot neither lying in a 3-ball inside the solid 
torus nor just wrapping one of the torus cycles) and then embedding the solid torus into the $S^3$ as another non-trivial knot.
\item Hyperbolic. In this case the compliment of the knot $S^3 \backslash \gamma$ is a hyperbolic space.
\item Torus. This family is very well-studied. Such knots are characterized by two numbers: $n$ and $m$. They can be obtained by wrapping the $S^1$ $n$ times over one cycle on a two-torus and $m$ times over the other cycle without 
self-intersections.
\end{itemize}
Obvious property of the torus knot is $K_{n,m}=K_{m,n}$. Actually, if $n$ and $m$ are not co-prime, it will be a link rather than a knot: 
link is a collection of knots which do not intersect. Also, $(n,1)$ and $(1,m)$ represent unknot. Therefore the most simple example is $(3,2)$ knot, so-called
trefoil knot(see Figure \ref{fg:kn32}).

The algebraic knot in $S^3$ which is the main
object in this section can be described by the intersection
of $S^3$ with some algebraic curve. If we realize the
sphere as
\beq
|z_1|^2 + |z_2|^2 =1
\eeq
then  the simplest $(p,q)$ torus knots which can be obtained
from the unknot by the $SL(2,Z)$ action corresponds
to the curve
\beq
z_1^p = z_2^q
\eeq
which is called Seifert surface of the knot. 
The sphere is invariant under
\beq
z_1\rightarrow e^{i \theta}z_1 \qquad z_2\rightarrow e^{i \theta}z_2
\eeq
while the Seifert surface under
\beq
z_1\rightarrow e^{iq\theta}z_1 \qquad z_2\rightarrow e^{ip\theta}z_2
\eeq

In the CS theory  the knot invariants one can evaluated from  the corresponding 
Wilson loop vacuum expectation value. The useful tool is the knot operator
introduced in \cite{labastida}
\beq
W_R^{n,m}|p>= \sum_{\mu \in M_R}exp[ -i\mu^2\frac{nm}{k+N} -
2\pi i \frac{m}{k+N}p\mu]|p+n\mu>
\eeq
where $M_R$ is the set of weights corresponding to the irreducible
representation R and $|p>$ is the element of the basis of the Hilbert space
of the SU(N) CS theory on the torus labeled by the weights p.
When evaluating the vev of Wilson loop one performs the
Heegaard cut of $S^3$ into two solid tori. Then the torus knot is introduced
on the surface of one of the solid tori by the action of the knot operator
on the corresponding vacuum state. In the standard framing the vev 
of Wilson loop is given by
\beq
<W^{n,m}_R> = E^{2\pi} \frac{<p|SW_R^{n,m}|p>}{<\rho|S|\rho>}
\eeq
where S - is the operator of S-transformation from $SL(2,Z)$.

\section{Appendix B. Higher (q,t) Catalan numbers}

In this Appendix we briefly review higher (q,t) deformed Catalan numbers  $C_n^k(q,t)$ which enter 
the expression for the $T_{n,nk+1}$  superpolynomial at $a=0$. There are several definitions 
of the higher Catalan numbers related to the geometry of the Hilbert schemes of points,
symmetric functions, representation theory and combinatorics of paths. To orient the reader
we provide a few of them
\begin{itemize}

\item   
Let us introduce the elementary symmetric functions $e_n$, Macdonald polynomial $H_{\mu}$ for the
partition $\mu$, the Hall product $<.,.>$ on the symmetric functions and $\Lambda$ - the ring 
of the symmetric functions. There is the so-called Nabla operator $\nabla$ which act on the Macdonald basis
as 
\beq
\nabla H_{\mu}= T_{\mu}H_{\mu},\qquad T_{\mu}=q^{n(\mu '}t^{n(\mu)}, \qquad 
n(\mu)= \sum_{x\in dg(\mu)}l(x)
\eeq
where $\mu '$ denotes the transpose of $\mu$. The  $C_n^k(q,t)$ in terms of the symmetric 
functions are defined as follows
\beq
C_n^k(q,t)= <\nabla^k(e_n),e_n>
\eeq

\item
One can define the higher Catalans in terms of the so-called diagonal harmonics. To this aim 
consider the polynomial ring $C(x_1,y_1,\dots, x_n,y_n)$  The symmetric group $S_n$ 
acts diagonally $\omega x_i = x_{\omega(i)}$ $\omega y_i = y_{\omega(i)}$ , $\omega \in S_n$.
Introduce the ideal generated by all alternating polynomials and let be $\mathfrak{m}$ the maximal ideal
generated by $x_1,y_i,\dots x_n,y_n$ . Let $M^k=I^k/\mathfrak{m}I^k$. It is possible do introduce
the double grading is the space of polynomials according the degrees in x and y variables. 
The grading tells that bi-degree $(d_1,d_2)$ corresponds to the situation when all monomials
of the polynomial have equal bi-degree $(d_1,d_2)$. The higher Catalans are now defined as
\beq
C_n^k(q,t)=\sum_l\sum_s q^l t^s dim M^k_{l,s}
\eeq
where  $M^k_{l,s}$ is the bihomogeneous component of $M^k$ of bidegree (l,s).

\item
The last definition which is the closest to our context is based on the 
Hilbert scheme of n points on $C^2$ $Hilb^n(C^2)$. Let us define
$O(k)= O(1)^{\otimes k}$. where $O(1) =det T$ and T is the tautological rank 
n bundle.  The double grading in this case is introduced on the set of
global section $H^0(Z_n,O(k))$ where  $Z_n$ tells that all points are sitting at the top of each other.
The $C_n^k(q,t)$ are now defined as follows
\beq
C_n^k(q,t) =\sum_l \sum_s q^l t^s dim H^0(Z_n,O(k))_{l,s}
\eeq

\end{itemize}

Let us emphasize that  the key property of the higher Catalans which is important 
in our study is that they provide the description of the properties of the  set of
points sitting at the top of each other.

\section{Appendix C. The Dunkl operators and rational DAHA}

In this Appendix we briefly describe the rational double affine Hecke algebras (DAHA) 
$H_c$ and their finite-dimensional representations relevant for the invariants of the torus knots.
The rational DAHA of type $A_{n-1}$ with parameter c is generated by the $V= C^{n-1}, V^*$
and the permutation group. $S_n$ with the following relations
\beq
\sigma x\sigma^{-1}=\sigma(x), \qquad \sigma y\sigma^{-1}=\sigma(y)
\eeq
\beq
x_1x_2= x_2x_1 \qquad y_1y_2= y_2y_1
\eeq
\beq 
yx -xy =<y,x> -c \sum_{s\in \mathfrak S}<\alpha_s,x><y,\alpha_s^{v}>s
\eeq
where $\mathfrak{S}$ is the set of all transpositions, and $\alpha_s,\alpha_s^{v}$
are the corresponding roots and coroots.

It is convenient to introduce the Dunkl operators
\beq
D_i =\frac{\partial}{\partial x_i} - c \sum_{i\neq j }\frac{s_{ij}-1}{x_i - x_j}
\eeq
and introduce the space of polynomial functions on V, where elements of V acts
by the multiplications and $V^*$ by the Dunkl operators. This representation is denoted by $M_c$.
It is known \cite{berest} that for rational $c=m/n$ DAHA has unique finite dimensional
representation $L_{m/n}$ which was identified as the factor $L_c=M_c/I_c$ where $I_c$
is the ideal generated by the following set of the homogeneous polynomials $f_i$  of degree m
\beq
f_i= Coef_m[(1-zx_i)^{-1}\prod_{i=1}^n (1 -zx_i)^{m/n}]
\eeq
They are annihilated by the Dunkl operators 
\beq
D_k(f_i)=0
\eeq
and therefore are  invariants under the DAHA action. The dimension of the finite-dimensional
representation is 
\beq
dim L^{m/n} = m^{n-1}
\eeq

There is $SL(2,R)$  subalgebra of DAHA which involves the Hamiltonian of the rational complexified Calogero model
\beq
H_{Cal}= \sum_i D_i^2
\eeq
the rest of the operators from this subalgebra are
\beq
K=\frac{1}{2}\sum_i(x_iD_i +D_ix_i)      \qquad    J_1= \sum_i  x_i^2
\eeq
Let us also note that there is the so-called shift operator which acts by changing $c\rightarrow c+1$ 
and is the counterpart of the Nabla operator in the theory of the symmetric functions and the cut- and-join
operator. It corresponds to the shift of the level of 5d CS term.

\section{Appendix D. The  Ruijsenaars-Shneyder  many-body integrable  model from the perturbed CS theory}

The  surface operator which survives at the road from UV to IR  \cite{rastelli} induces
the action of the quantum trigonometric RS Hamiltonian
which is simply described in terms o the perturbed 3d CS theory \cite{gn95}.
The phase space in this model  is identified with the space of flat connections 
on the torus with the marked point and particular monodromy around
while the Hamiltonian can be described as the Wilson loop around one cycle.

Equivalently is can be obtained via the Hamiltonian reduction.
To perform the Hamiltonian reduction  replace the space of
two dimensional gauge fields by the
cotangent space to the 
loop group:
$$
T^{*} \hat G   = \{ ( g(x), k_{x} + P(x) ) \}
$$
The relation to the two dimensional
construction
is the following. Choose a non-contractible circle $\bf S^{1}$
on the two-torus
which does not pass through the marked point $p$. Let $x,y$
be the coordinates on the torus
and $y=0$ is the equation of the $\bf S^{1}$.
The periodicity  of $x$ is $\beta$ and that of $y$ is $R$.
Then
$$
P(x) = A_{x}(x,0),
g(x) =P\exp\int_{0}^{R} A_{y}(x,y) dy.
$$
The moment map equation looks as follows:
\beq
k g^{-1}_{x}
g + g^{-1}Pg - P = J \delta(x),
\eeq
with $k = {1\over{R\beta}}$. The solution of
this equation in the gauge $P = {\rm diag}(q_{1}, \ldots, q_{N})$
leads to
the Lax operator $A = g(0)$ with $R,\beta$
exchanged. On the other hand,
if we  diagonalize
$g(x)$:
\beq
g(x) =  {\rm diag} \left( z_{1} =
e^{{\sqrt -1}R q_{1}}, \ldots, z_{N} = e^{{\sqrt -1}R q_{N}} \right)
\eeq
then a similar calculation
leads to the Lax operator
$$
B = P\exp\oint{1\over{k}} P(x)dx =  {\rm diag} (
e^{{\sqrt -1}\theta_{i} } ) \exp {\sqrt -1}R\beta\nu {\rm r}
$$
with
$$
{\rm r}_{ij} =
{1\over{1- e^{{\sqrt -1}Rq_{ji}}}}, i\neq j; \quad {\rm r}_{ii} = - \sum_{j\neq i}
{\rm r}_{ij}
$$
thereby establishing the duality $A \leftrightarrow B$
explicitly.

\printbibliography

\end{document}